\documentclass[10pt]{iopart}
\pdfoutput=1

\usepackage{graphicx}
\usepackage{iopams}  
\usepackage{color}
\usepackage[dvipsnames]{xcolor}
\usepackage{hyperref}

\begin{document}

\title{Bootstrap and diffusion percolation transitions in three-dimensional lattices}

\date{\today}

\author{Jeong-Ok Choi$^1$ and Unjong Yu$^2$}
\address{$^1$ Division of Liberal Arts and Sciences, Gwangju Institute of Science and Technology, Gwangju 61005, South Korea}
\address{$^2$ Department of Physics and Photon Science, Gwangju Institute of Science and Technology, Gwangju 61005, South Korea}
\ead{uyu@gist.ac.kr}

\begin{abstract}
We study the bootstrap and diffusion percolation models in the simple-cubic (sc), body-centered cubic (bcc), and face-centered cubic (fcc) lattices using the Newman-Ziff algorithm. The percolation threshold and critical exponents were calculated through finite-size scaling with high precision in the three lattices.
In addition to the continuous and first-order percolation transitions, we found a double transition, which is a continuous transition followed by a discontinuity of the order parameter.
We show that the continuous transitions of the bootstrap and diffusion percolation models have the same critical exponents as the classical percolation within error bars and they all belong to the same universality class.
\end{abstract}



\section{Introduction}
The bootstrap percolation (BP) \cite{Chalupa79} and diffusion percolation (DP) \cite{Adler88} models have attracted continuous attention \cite{Adler91,DeGregorio09} since they have close relation with disordered dilute magnetic systems \cite{Chalupa79,Adler87,Sabhapandit02}, neuronal activity \cite{Soriano08,Amini10}, jamming transition \cite{Gregorio05,Toninelli06}, disease spreading \cite{Balogh98}, failure of computer storage arrays \cite{Kirkpatrick02}, diffusion of innovations \cite{Lelarge08,Helbing12}, and zero-temperature dynamics of the Ising model \cite{Fontes02,Yu17}.
The DP and BP processes start from a lattice whose sites are occupied randomly with the probability $p$. 
In the BP model, occupied sites that have less than $m$ occupied neighbors become empty recursively until all the occupied sites have at least $m$ occupied neighbors.
In the DP, empty sites that have at least $k$ occupied neighbors become occupied recursively until all the empty sites have less than $k$ occupied neighbors.
Although the DP is sometimes called also as the BP \cite{Baxter10,Gravner12,Gao15},
we differentiate them in this paper.

There have been many studies of the BP and DP on two-dimensional lattices \cite{Kogut81,Branco84,Khan85,Branco86,Adler88,Chaves95,Medeiros97,Gravner12,Choi19}.
According to our recent study on the eleven Archimedean lattices \cite{Choi19}, 
the BP with $m\leq m_c$ and the DP with $k\geq (\Delta+1-m_c)$ have
percolation phase transitions at finite initial occupation probability $p$ with $0<p<1$, where $\Delta$ is the coordination number of the lattice.
The percolation thresholds ($p_{\mathrm{\scriptscriptstyle BP(}m\scriptscriptstyle )}$ and $p_{\mathrm{\scriptscriptstyle DP(}k\scriptscriptstyle )}$) depend on the parameters $m$ and $k$, respectively.
The percolation transitions are continuous and they belong to the same universality class as the classical percolation (CP) transition in two-dimensional lattices. 
In Ref.~\cite{Choi19}, we found that $m_c=\lfloor (\Delta+1)/2 \rfloor$ except for the bounce lattice, which is also known as the (3,4,6,4) lattice; in fact for the bounce lattice $m_c=\lfloor (\Delta+1)/2 \rfloor+1=3$ with $\Delta=4$.
The BP with $m > m_c$ and the DP with $k < (\Delta+1-m_c)$ have
first-order percolation transitions with the percolation thresholds 
$p_{\mathrm{\scriptscriptstyle BP(}m\scriptscriptstyle )}=1$ and $p_{\mathrm{\scriptscriptstyle DP(}k\scriptscriptstyle )}=0$, respectively.

As for three-dimensional lattices, previous studies have been mainly focused on the BP in the simple-cubic (sc) lattice \cite{Kogut81,Manna89,Adler90,Branco99,Kurtsiefer03}. 
The BP has continuous percolation transitions for $m\leq3$ and first-order percolation transitions with $p_{\mathrm{\scriptscriptstyle BP(}m\scriptscriptstyle )}=1$ for $m\geq4$ in the sc lattice \cite{Aizenman88,Cerf99,Balogh09}.
There exists a debate about the universality class of the percolation transition for the BP with $m=3$.
Some people insist that the universality class is different from the CP as they estimated the value of the critical exponent to be $\beta=0.82(4)$ in Ref.~\cite{Kogut81} and  $\beta=0.6(1)$ in Ref.~\cite{Adler90}, which is different from that of the CP.
(Recent estimate of the critical exponent of the CP in the sc lattice is $\beta=0.4180(6)$ \cite{Xu14}.)
To the contrary, Ref.~\cite{Branco99} reported $\beta=0.37(3)$ and argued they belong to the same universality class.
Due to low precision of the results, the universality class of the BP in three-dimensional lattices is far from conclusive.

In this paper, the BP and DP models are studied by the Newman-Ziff algorithm in three kinds of three-dimensional lattices: simple-cubic (sc), body-centered cubic (bcc), and face-centered cubic (fcc) lattices.
We classify the percolation transitions into first-order and continuous transitions.
In addition, we found a double transition, which is a continuous percolation transition followed by a discontinuity of the order parameter, in the fcc lattice.
This kind of double transition is observed for the DP in complex networks \cite{Baxter10,Wu14,Gao15}, 
but has not been reported before in lattices, as far as we know.
We calculated the percolation threshold and critical exponents ($\nu$ and $\beta$) for continuous transitions with high precision to show that the BP and DP belong to the same universality class as the CP in three dimensions.

\section{Methods}

In order to investigate the BP and DP with high precision, the Newman-Ziff algorithm \cite{Newman00} was adopted. Contrary to the conventional method that calculates physical quantity $\langle Q(p) \rangle$ at fixed occupation probability $p$, in the Newman-Ziff method $\langle Q(n) \rangle$ is calculated as a function of number of occupied sites $n$ and then $\langle Q(p) \rangle$ is obtained by a convolution with the binomial distribution
\begin{eqnarray}
\langle Q(p) \rangle = \sum_{n=0}^{N} \frac{N!}{n! (N-n)!}
                       p^n (1-p)^{N-n}
                       \; \langle Q(n) \rangle , \label{caninical_transf}
\end{eqnarray}
where $N$ is the total number of sites of the lattice. 
Therefore, once $\langle Q(n) \rangle$ is obtained precisely, $\langle Q(p) \rangle$ and its derivatives for any $p$ can be calculated just by equation~(\ref{caninical_transf}) \cite{Newman01,Martins03,Choi19}.
This method was recently extended to study the DP and BP, and was applied to the two-dimensional Archimedean lattices \cite{Choi19}.

The new algorithms made the calculation of the DP and BP as efficient as the CP except for the BP with $m\geq3$, where it takes more than ten times than the CP \cite{Choi19}. 
Since the lattices studied in this paper have large coordination numbers, we propose another algorithm for the BP, which is more complicated but far faster than the algorithm of Ref.~\cite{Choi19} for $m\geq3$. 
It uses the close relation between the DP of $k$ and BP of $m=\Delta+1-k$ in $\Delta$-regular lattices, where every site has the same coordination number $\Delta$. 
In the Newman-Ziff method, the DP and BP processes begin from $\{ S_0, S_1, \cdots, S_{N-1}, S_N \}$ and $\{ S'_0, S'_1, \cdots, S'_{N-1}, S'_N \}$, respectively.
States $S_n$ and $S'_n$ represent initial states with $n$ occupied sites chosen at random. 
And then, the diffusion and culling processes are performed from $S_n$ in the DP and from $S'_n$ in the BP, respectively. 
Interestingly, if we consider occupied (resp. empty) sites in the DP as the empty (resp. occupied) sites in the BP, the diffusion process from $S_n$ of the DP is the same as the culling process of the BP from $S'_{N-n}$ if $k+m = \Delta+1$. 
Therefore, without loss of generality, the DP processes of $k$ from the sequence in the reversed order, say $\{S_N, S_{N-1},  \cdots, S_1, S_0\}$, can be considered as the BP processes of $m=\Delta+1-k$ in the sequence of $\{S'_0, S'_1, \cdots, S'_{N-1}, S'_N\}$. 
As a result, whenever we sample a sequence for the DP model, we can use the sequence for the BP by reversing the order. The new algorithm to calculate quantities as a function of initial filling for the DP  ($Q_{\mathrm{\scriptscriptstyle DP}}(n_{\mathrm{\scriptscriptstyle DP}})$) and for the BP ($Q_{\mathrm{\scriptscriptstyle BP}}(n_{\mathrm{\scriptscriptstyle BP}})$)
is summarized as follows.
\\ \\
(1) Initially, all sites are empty.\\
(2) Make an array of all the sites in a random order.\\
(3) Initialize the step number to be $n_{\mathrm{\scriptscriptstyle DP}}=1$. \\
(4-a) If the $n_{\mathrm{\scriptscriptstyle DP}}$th site of the array made in step (2) is empty, fill it.
    If any other site has at least $k$ occupied neighbors, fill it recursively.
    Push the newly filled sites in the stack with the current step number $n_{\mathrm{\scriptscriptstyle DP}}$. \\
(4-b) Calculate $Q_{\mathrm{\scriptscriptstyle DP}}(n_{\mathrm{\scriptscriptstyle DP}})$ for the DP of $k$. \\
(4-c) Increase $n_{\mathrm{\scriptscriptstyle DP}}$ by one. \\
(5) Repeat step (4), until all the sites are occupied. \\ \\
(6) All sites are set to be empty.\\
(7) The site that is popped from the stack is occupied. 
   Repeat this until all the sites of the same $n_{\mathrm{\scriptscriptstyle DP}}$ are popped.
   Calculate $Q_{\mathrm{\scriptscriptstyle BP}}(n_{\mathrm{\scriptscriptstyle BP}})$ for the BP of $m=\Delta+1-k$ with $n_{\mathrm{\scriptscriptstyle BP}}=N-n_{\mathrm{\scriptscriptstyle DP}}$. \\
(8) Repeat step (7) until the stack is empty. \\ \\
The whole steps are repeated to average $Q_{\mathrm{\scriptscriptstyle DP}}(n_{\mathrm{\scriptscriptstyle DP}})$ and $Q_{\mathrm{\scriptscriptstyle BP}}(n_{\mathrm{\scriptscriptstyle BP}})$.
From step (1) to step (5), data for the DP of $k$ are gathered and the DP process is saved in the stack, which is a linear data structure in the last-in-first-out (LIFO) order.
The stack is used for the BP of $m=\Delta+1-k$. 
Therefore, by one simulation, both the DP and BP results with $m+k=\Delta+1$ are obtained. 
Note that this technique can be used in any dimensions including complex networks if it is a regular graph.
This algorithm gives mathematically the same results as those in Ref.~\cite{Choi19}.
In the case of irregular networks, for a given BP with $m$, a formula for $k$ in the DP does not exist because each site has a different coordination number. 
Therefore, this algorithm cannot be used for an irregular network in the present form.

Using this algorithm, we calculated the strength of the largest cluster
(the probability that a site belongs to the largest cluster; $P_{\infty}$),
percolation probability in any one direction ($P_{w1}$), and in any two directions ($P_{w2}$), and in all the three directions ($P_{w3}$) as a function of $p$ for the CP, BP, and DP. Average cluster size excluding the largest one also has maximum at the percolation threshold, but it was not used in this study because of large deviation from finite-size scaling \cite{Choi19}.

In this work, we consider three kinds of three-dimensional lattices (sc, bcc, and fcc lattices) with periodic boundary condition, where the percolation is defined by the existence of a wrapping cluster.
The number of lattice sites is $N=L^3$, where linear size $L$ is from $22$ to $240$.

The percolation thresholds ($p_{\mathrm{\scriptscriptstyle DP(}k\scriptscriptstyle )}$ and $p_{\mathrm{\scriptscriptstyle BP(}m\scriptscriptstyle )}$) 
of infinite lattices
are determined by finite-size scaling.
The percolation threshold estimate of a finite lattice $p_c(L)$ is
determined by the probabilities of initial filling ($p$) 
that make some physical quantities maximum.
The percolation threshold can also be found by the crossing points 
of percolation probabilities ($P_{w1}$, $P_{w2}$, and $P_{w3}$) of different lattice sizes in continuous percolation transitions \cite{Newman01}.
We averaged the percolation threshold values obtained 
by these methods to get the final estimate.
Critical exponents $\nu$ and $\beta$ are obtained by the derivative of
percolation probability and strength of the largest cluster, respectively \cite{Lobb80,Martins03}.
We ignore the correction-to-scaling \cite{Ballesteros99,Ziff11} in the calculation of the percolation threshold and critical exponents, since its effect is estimated to be small.

\begin{figure*}[!b]
\centering
\includegraphics[width=\columnwidth]{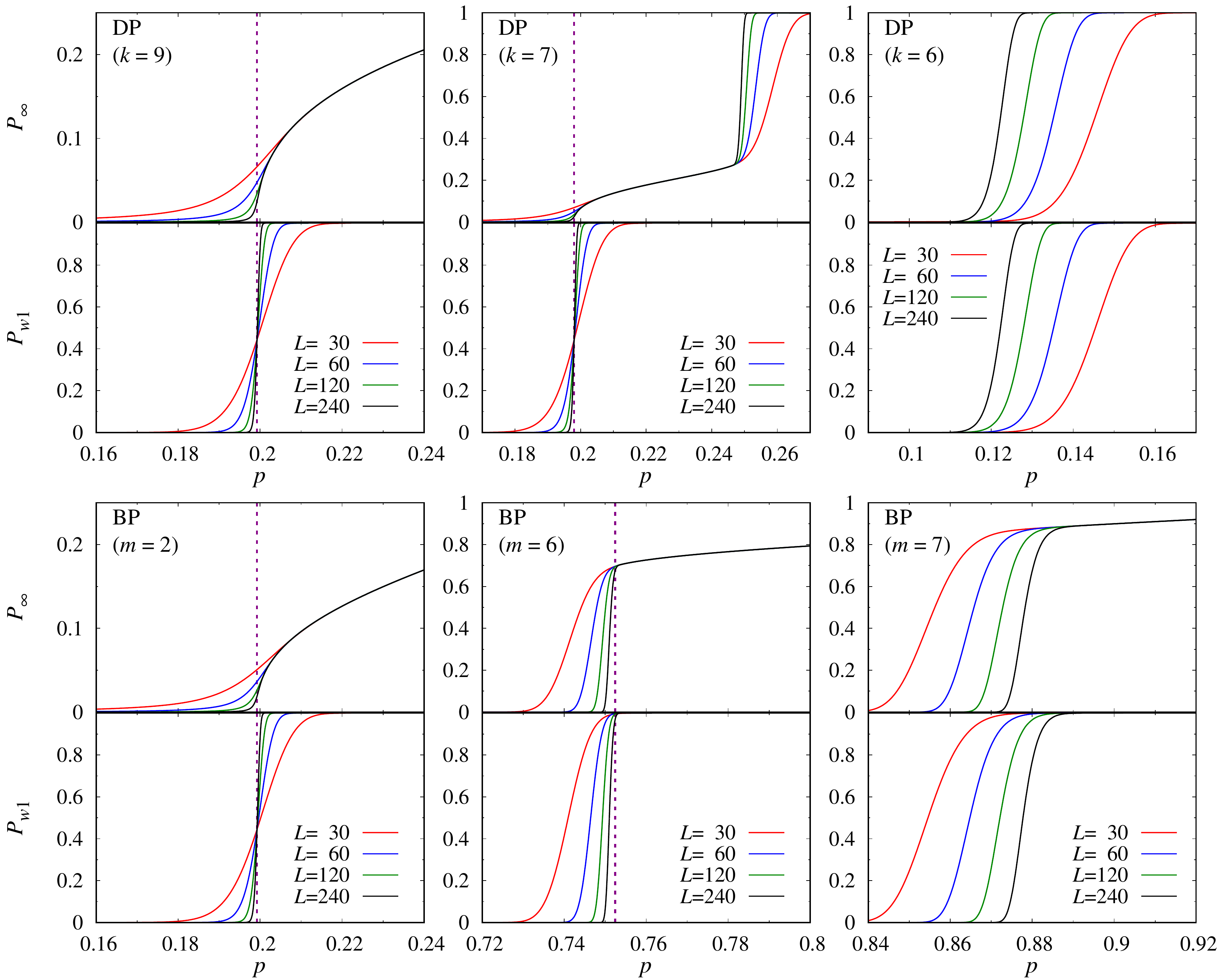}
\caption{Strength of the largest cluster ($P_{\infty}$) and
percolation probability in any direction ($P_{w1}$) as a function of initial filling probability ($p$) 
for the DP and BP in the fcc lattice for various linear sizes $L$. 
Vertical dotted lines indicate our estimates of percolation thresholds
($p_{\mathrm{\scriptscriptstyle DP(}k\scriptscriptstyle )}$ 
and $p_{\mathrm{\scriptscriptstyle BP(}m\scriptscriptstyle )}$).
Note that the BP of $m=6$ have first-order percolation transition at $p_{\mathrm{\scriptscriptstyle BP(}m\scriptscriptstyle )}<1$, and the DP of $k=6$ and the BP of $m=7$ have first-order percolation transitions at $p_{\mathrm{\scriptscriptstyle DP(}k=6\scriptscriptstyle )}=0$ and $p_{\mathrm{\scriptscriptstyle BP(}m=7\scriptscriptstyle )}=1$, respectively.}
\label{P_vs_p}
\end{figure*}

\begin{figure*}[!tb]
\centering
\includegraphics[width=\columnwidth]{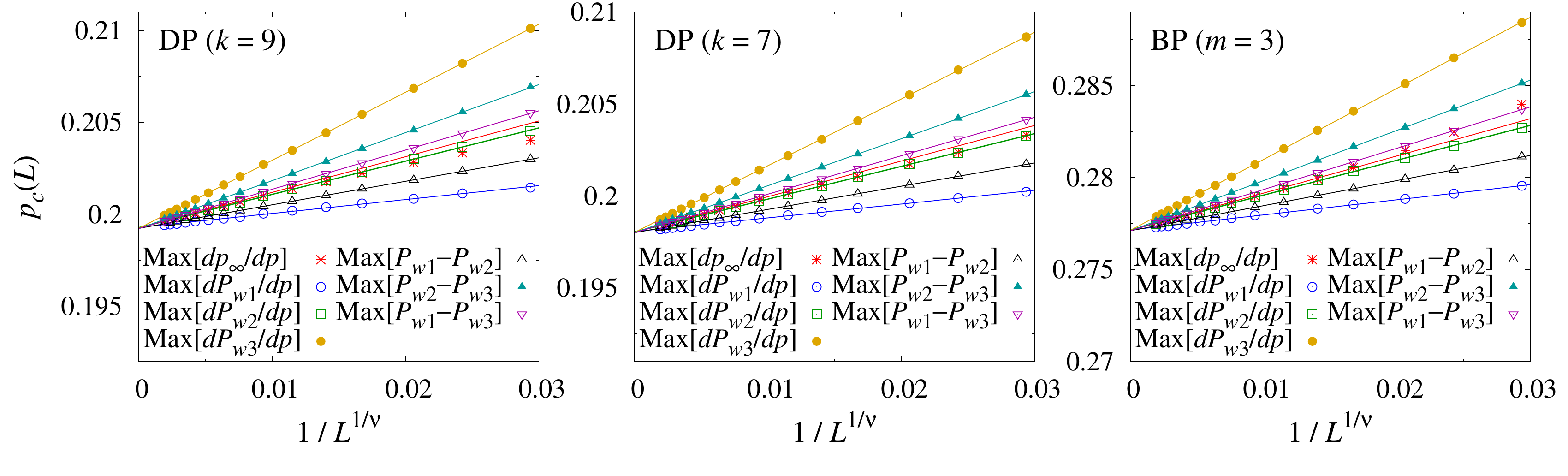}
\caption{The percolation threshold estimate of finite systems $p_c(L)$ 
as a function of linear size $L$ 
for the DP ($k=9, 7$) and BP ($m=3$) in the fcc lattice.
They were determined by the probabilities of initial filling ($p$) that give the maximum of 
$dP_{\infty}/dp$, $dP_{w1}/dp$, $dP_{w2}/dp$, $dP_{w3}/dp$, $(P_{w1}-P_{w2})$, $(P_{w2}-P_{w3})$, and $(P_{w1}-P_{w3})$.
Solid lines are from fitting of $[p_c(L) - p_c^{(\infty)}] \sim L^{-1/\nu}$.
For $\mathrm{Max}[dP_{\infty}/dp]$, small systems ($L<100$) were excluded in the fitting.}
\label{pc_vs_L}
\end{figure*}

\begin{figure*}[!tb]
\centering
\includegraphics[width=\columnwidth]{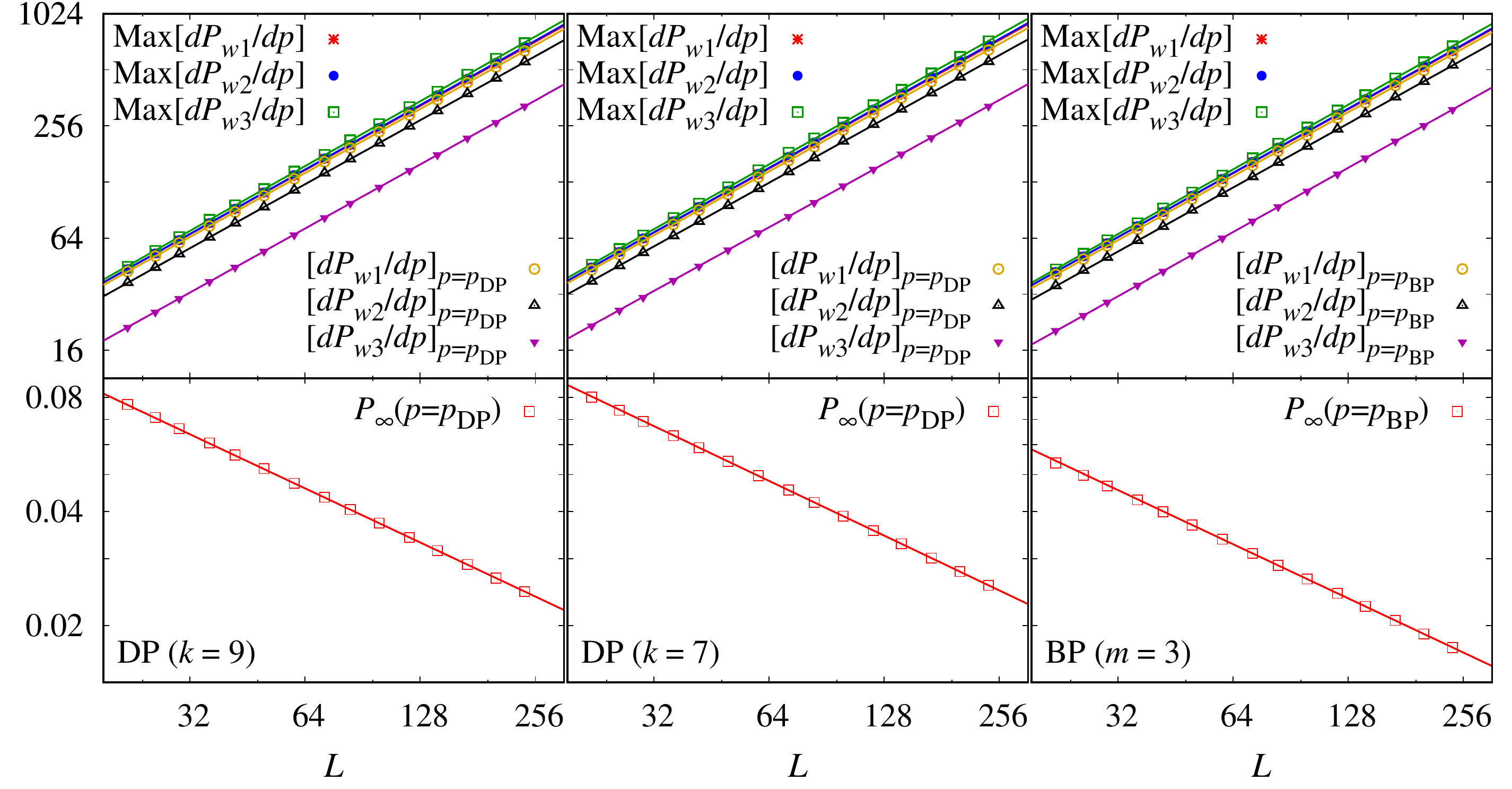}
\caption{Maximum values of $dP_{w}/dp$ and values of $dP_{w}/dp$ at the percolation
threshold $p_c^{(\infty)}$ in the upper panels, and
values of $P_{\infty}$ at the percolation
threshold $p_c^{(\infty)}$ in the lower panels as a function of
system's linear size $L$ in the fcc lattice in log-log scale. 
$P_{w}$ stands for $P_{w1}$, $P_{w2}$, and $P_{w3}$.
Solid lines are from fitting of equations~(\protect\ref{scale2})-(\protect\ref{scale4}).}
\label{crit_exp}
\end{figure*}

\begin{table*}[!tb]
\caption{\label{table1} Name, coordination number ($\Delta$), 
percolation threshold ($p_{\mathrm{\scriptscriptstyle BP(}m\scriptscriptstyle )}$ 
and $p_{\mathrm{\scriptscriptstyle DP(}k\scriptscriptstyle )}$), and critical exponents ($\nu$ and $\beta$) for the CP, BP, and DP with the percolation transition of three three-dimensional lattices. (The rows of $m=0$ in the BP represent the results of the CP.)
Note that the percolation transitions in this table are continuous except for the BP with $m=6$ in the fcc lattice.}
\centering
\begin{tabular}{|c|cccc|cccc|}
\hline
       & \multicolumn{4}{|c|}{Bootstrap percolation} & \multicolumn{4}{|c|}{Diffusion percolation} \\
        \cline{2-9}
Lattice & $m$ & $p_{\mathrm{\scriptscriptstyle BP(}m\scriptscriptstyle )}$  & $\nu$  & $\beta$ & $k$ & $p_{\mathrm{\scriptscriptstyle DP(}k\scriptscriptstyle )}$
        & $\nu$  & $\beta$ \\ \hline
sc           & $0$ & 0.31162(1) & 0.872(5) & 0.415(3) &     &            &          &          \\
($\Delta\!=\!6$) & $1$ & 0.31161(1) & 0.875(3) & 0.417(2) & $6$ & 0.31023(1) & 0.876(3) & 0.419(2) \\
             & $2$ & 0.31160(1) & 0.878(4) & 0.419(3) & $5$ & 0.29354(1) & 0.875(3) & 0.418(3) \\
             & $3$ & 0.57256(1) & 0.878(4) & 0.418(5) & $4$ & 0.23732(1) & 0.877(3) & 0.419(4) \\
             \hline
bcc          & $0$ & 0.24597(2) & 0.875(3) & 0.417(2) &     &            &          &          \\
($\Delta\!=\!8$) & $1$ & 0.24596(1) & 0.874(3) & 0.417(3) & $8$ & 0.24594(1) & 0.876(3) & 0.419(3) \\
             & $2$ & 0.24597(1) & 0.874(3) & 0.416(3) & $7$ & 0.24536(1) & 0.875(4) & 0.418(3) \\
             & $3$ & 0.41677(1) & 0.877(3) & 0.415(5) & $6$ & 0.24000(1) & 0.873(3) & 0.416(2) \\
             & $4$ & 0.67258(1) & 0.879(4) & 0.411(9) & $5$ & 0.21755(1) & 0.876(3) & 0.418(2) \\
             \hline
fcc          & $0$ & 0.19923(1) & 0.874(5) & 0.419(3) &     &            &          &          \\
($\Delta\!\!=\!\!12$)& $1$ & 0.19924(1) & 0.874(4) & 0.417(3) & $12$& 0.19924(1) & 0.874(4) & 0.415(3) \\
             & $2$ & 0.19923(1) & 0.878(4) & 0.420(3) & $11$& 0.19924(1) & 0.876(4) & 0.418(3) \\
             & $3$ & 0.27712(1) & 0.874(5) & 0.416(4) & $10$& 0.19924(1) & 0.875(3) & 0.418(2) \\
             & $4$ & 0.40625(1) & 0.878(5) & 0.418(4) & $9$ & 0.19924(1) & 0.875(4) & 0.417(3) \\
             & $5$ & 0.57216(1) & 0.881(6) & 0.416(5) & $8$ & 0.19922(1) & 0.875(3) & 0.418(3) \\
             & $6$ & 0.75243(6) & -        & -        & $7$ & 0.19800(1) & 0.879(5) & 0.421(4) \\
             \hline
\end{tabular}
\end{table*}

\section{Results}

Figure~\ref{P_vs_p} shows the strength of the largest cluster $P_{\infty}(p,L)$
and percolation probability in any direction $P_{w1}(p,L)$ for the DP ($k=9,7,6$) and BP ($m=2,6,7$) in the fcc lattice.
The other percolation probabilities, $P_{w2}(p,L)$ and $P_{w3}(p,L)$, also show steep increase at $p_{c}(L)$.
In all the cases of the three lattices (sc, bcc, and fcc lattices), we found $m_c=\lfloor (\Delta+1)/2 \rfloor$ like many two-dimensional lattices:
The BP with $m\leq m_c$ and the DP with $k\geq (\Delta+1-m_c)$ have
percolation phase transitions at finite $p$ of $0<p<1$, and 
the BP with $m > m_c$ and the DP with $k < (\Delta+1-m_c)$ have 
first-order percolation transitions with the percolation thresholds 
$p_{\mathrm{\scriptscriptstyle BP(}m\scriptscriptstyle )}=1$ and $p_{\mathrm{\scriptscriptstyle DP(}k\scriptscriptstyle )}=0$, respectively.
In the case of the sc lattice, it was rigorously proven that the BP with $m=m_c+1=4$ has $p_{\mathrm{\scriptscriptstyle BP(}4\scriptscriptstyle )}=1$ \cite{Schonmann92} and $p_c(L)$ approaches 1 very slowly ($[1-p_{c}(L)] \sim 1/\log(\log(L))$) \cite{Cerf99};
for the DP with $k=(\Delta-m_c)=3$, it was also proven that $p_{c}(L) \sim 1/\log(\log(L))$ \cite{Balogh09}.
Such a behavior is observed in the BP with $m=m_c+1$ and the DP with $k=(\Delta-m_c)$ in those three lattices.
For the fcc lattice (the BP with $m=7$ and the DP with $k=6$), see figure~\ref{P_vs_p}.
Note that all the percolation phase transitions of the BP with $m\leq m_c$ and the DP with $k\geq (\Delta+1-m_c)$ are continuous except the BP with $m =m_c=6$ in the fcc lattice;
the transition is of first-order and the percolation threshold $p_{\mathrm{\scriptscriptstyle BP(}m\scriptscriptstyle )}$ is less than 1.
Interestingly, we observe a double transition in the DP with $k=\Delta+1-m_c=7$ in the fcc lattice.
The double percolation transition is a continuous percolation transition at the percolation threshold and a discontinuous jump of the order parameter ($P_{\infty}$) at higher value of initial occupation probability $p$.
This kind of double percolation transition has been observed only in the DP of complex networks (Erd\H{o}s-R\'{e}nyi random networks with and without community structures \cite{Baxter10,Wu14} and regular random networks \cite{Gao15}). 

To study the condition for the existence of the double transition, we investigated the DP in two-, three-, and four-dimensional hypercubic lattices with various coordination numbers. 
In two dimensions, four kinds of lattices with $\Delta=12$, 20, 24, and 28 were made by including up to sixth nearest neighbors in the square lattice and they were examined; we found double transitions for $\Delta=24$ and $28$. 
In three dimensions, the double transition is observed in the sc lattice with nearest- and next-nearest-neighbors ($\Delta=18$) like the fcc lattice ($\Delta=12$). 
In the four-dimensional hypercubic lattice with only nearest-neighbors ($\Delta=8$), we found a double transition too. 
Therefore, we conclude that a double transition exists in lattices with large coordination number, and the minimum coordination number for a double transition to appear decreases rapidly with spatial dimension.


The percolation threshold of infinite lattices [$p_c^{(\infty)}=\lim_{L\rightarrow\infty}p_c(L)$] is determined by finite-size scaling: $[p_c(L) - p_c^{(\infty)}] \sim L^{-1/\nu}$ in three-dimensional continuous phase transitions. 
The percolation threshold estimate for a finite lattice $p_c(L)$ is determined by the initial filling probabilities ($p$) that make the values of $\mathrm{d}P_{\infty}(p,L)/\mathrm{d}p$, $\mathrm{d}P_{w1}(p,L)/\mathrm{d}p$, $\mathrm{d}P_{w2}(p,L)/\mathrm{d}p$, $\mathrm{d}P_{w2}(p,L)/\mathrm{d}p$, $[P_{w1}(p,L)-P_{w2}(p,L)]$, $[P_{w2}(p,L)-P_{w3}(p,L)]$, and $[P_{w1}(p,L)-P_{w3}(p,L)]$ maximum.
Here, $[P_{w1}(p,L)-P_{w2}(p,L)]$ means the probability of percolation only in one direction and not the other two directions.
Figure~\ref{pc_vs_L} confirms the scaling behavior very well.
In the case of $\mathrm{d}P_{\infty}(p,L)/\mathrm{d}p$, however, deviation from the scaling is large in small lattices  
and so results of lattices of $L<100$ were excluded in the fitting. 
This kind of deviation in $\mathrm{d}P_{\infty}(p,L)/\mathrm{d}p$ exists in the other lattices too.
We also tried a fitting that includes the correction-to-scaling: $[p_c(L) - p_c^{(\infty)}] \sim L^{-1/\nu}(1+b L^{-\omega})$ with the scaling correction exponent $\omega=1.61(5)$ and a fitting parameter $b$ \cite{Lorenz98}.
The correction affects very little on the values of the percolation threshold: it makes the percolation threshold systematically smaller by about 0.005\% on average.
Since it is not clear whether the BP and the DP have the same value of the exponent $\omega$,
we did not include the effect of nonzero $\omega$ in the final results of this work.
The precision of this work is not enough to obtain a reliable value of the exponent $\omega$.
The percolation threshold of continuous percolation transition can also be found by the value of $p$ at the crossing points of $P_{w1}(p,L)$, $P_{w2}(p,L)$, and $P_{w3}(p,L)$ with various linear sizes ($L$) \cite{Newman01}, as shown in figure~\ref{P_vs_p}.
The two methods give consistent results; we found that the latter method gives smaller uncertainty.
The final estimate of the percolation threshold was obtained by taking the average.
Table~\ref{table1} shows the percolation threshold of all continuous transitions and the first-order transition for the BP with $m=6$ in the fcc lattice. 
We confirmed that the values of the percolation threshold for the CP are consistent with the most precise numerical results ($p_c^{(\infty)} = 0.31160768(15)$, $p_c^{(\infty)}=0.2459615(2)$, and $p_c^{(\infty)}=0.19923517(20)$ for the sc, bcc, and fcc lattices, respectively \cite{Xu14}) within relative errors of 0.003\%.
The percolation threshold results of the BP with $m=1$ or $m=2$ are the same as that of the CP within error bars; it is because the culling process in the BP with $m\leq2$ just removes isolated occupied sites and dangling ends (leaves) of percolation clusters \cite{Adler91}.
The percolation threshold of the BP with $m=3$ for the sc lattice is also
consistent with references ($p_{\mathrm{\scriptscriptstyle BP(}m=3\scriptscriptstyle )}=0.568(2)$ \cite{Kogut81}, $p_{\mathrm{\scriptscriptstyle BP(}m=3\scriptscriptstyle )}=0.5717(5)$ \cite{Adler90},  $p_{\mathrm{\scriptscriptstyle BP(}m=3\scriptscriptstyle )}=0.57256(6)$ \cite{Branco99}, and $p_{\mathrm{\scriptscriptstyle BP(}m=3\scriptscriptstyle )}=0.5726(1)$ \cite{Kurtsiefer03}), but our result has much smaller uncertainty.

The derivative of percolation probability and the strength of the largest cluster
 satisfy the scaling relations \cite{Lobb80,Martins03,Choi19}: 
\begin{eqnarray}
\mathrm{Max}\left[\mathrm{d}P_{w}(p,L)/\mathrm{d}p\right] \sim L^{1/\nu}, \label{scale2} \\
\left[\mathrm{d}P_{w}(p,L)/\mathrm{d}p\right]_{p=p_c^{(\infty)}} \sim L^{1/\nu}, \label{scale3} \\
\mbox{and } \; P_{\infty}(p_c^{(\infty)},L) \sim L^{-\beta/\nu} , \label{scale4}
\end{eqnarray}
where $P_{w}(p,L)$ stands for $P_{w1}(p,L)$, $P_{w2}(p,L)$, and $P_{w3}(p,L)$.
Therefore, the critical exponents $\nu$ and $\beta$ can be obtained by fitting.
Figure~\ref{crit_exp} shows that these values follow the scaling relation very well for the DP ($k=9,7$) and BP ($m=3$) in the fcc lattice.
The DP and BP with different values of $k$ and $m$ in the other kinds of lattice show equivalent behavior close to the continuous transition.
Table~\ref{table1} summarizes the critical exponents obtained in this work.
They are consistent with one another and reference values of the three-dimensional classical percolation model
($\nu=0.8765(18)$ and $\beta=0.4181(8)$ \cite{Ballesteros99}; $\nu=0.87619(12)$ and $\beta=0.4180(6)$ \cite{Xu14}) within error bars.
Therefore, we are quite convinced that continuous transitions of the BP and DP have the same critical exponents as the CP in three dimensions.
Another important quantity is the universal wrapping probability of the critical percolation in the thermodynamic limit $L \rightarrow\infty$: the values of $P_{w1}$, $P_{w2}$, and $P_{w3}$ at the percolation threshold $p_c^{(\infty)}$ \cite{Wang13}.
We found they are universal and do not depend on the type of percolation.
They were determined to be $P_{w1}(p_c^{(\infty)})=0.460(2)$, $P_{w2}(p_c^{(\infty)})=0.232(2)$, $P_{w3}(p_c^{(\infty)})=0.080(1)$ for the sc lattice, $P_{w1}(p_c^{(\infty)})=0.443(2)$, $P_{w2}(p_c^{(\infty)})=0.353(2)$, $P_{w3}(p_c^{(\infty)})=0.208(1)$ for the bcc lattice, and $P_{w1}(p_c^{(\infty)})=0.442(4)$, $P_{w2}(p_c^{(\infty)})=0.316(3)$, $P_{w3}(p_c^{(\infty)})=0.118(2)$ for the fcc lattice. The values for the sc lattice is consistent with previous results ($P_{w1}(p_c^{(\infty)})=0.46002(2)$ and $P_{w3}(p_c^{(\infty)})=0.08046(4)$ \cite{Wang13,Xu14}).
Note that the universal wrapping probability does depend on the type of the lattice \cite{Pinson94}; the lattices we used in this work are cubic for the sc lattice and rhombohedral for the bcc and fcc lattices \cite{Yu15}.
It is worth mentioning that the leaf-excluded bond percolation model was recently shown to belong to the universality class of the CP too \cite{Zhou15}.

\section{Summary}

We proposed a very efficient algorithm that can calculate the BP and the DP models simultaneously within the Newman-Ziff algorithm for regular lattices. 
This algorithm can be used in any network that is regular.
Using this algorithm and finite-size scaling, we studied the BP and DP in three three-dimensional lattices (sc, bcc, and fcc lattices) with high precision to calculate the percolation threshold and critical exponents $\nu$ and $\beta$.
We found that $m_c=\lfloor (\Delta+1)/2 \rfloor$ for the three three-dimensional lattices:
The BP with $m > m_c$ and the DP with $k < (\Delta+1-m_c)$ have first-order percolation transition with percolation threshold 1 and 0, respectively, for infinite-size lattices.
The BP with $m \leq m_c$ and the DP with $k \geq (\Delta+1-m_c)$ have continuous transitions for the sc and bcc lattices.
As for the fcc lattice, the BP with $m=m_c=6$ has a first-order percolation transition at finite $p_c^{(\infty)}$, and the DP with $k=(\Delta+1-m_c)=7$ has a double transition, which is a continuous percolation transition followed by a discontinuous jump of the order parameter at higher initial occupation probability. 
We found that the double percolation transition in the DP appears in lattices only when the coordination number is larger than the critical value $\Delta_{\mathrm{double}}$ and the value of $\Delta_{\mathrm{double}}$ decreases rapidly with increasing spatial dimension.
This is why the double percolation transition is rare in low-dimensional lattices while it is observed often in complex networks.
When the percolation transition is continuous, the critical exponents for the BP and DP are the same as the CP within error bars, and we conclude that the three percolation models all belong to the same universality class.

\ack
This work was supported by GIST Research Institute (GRI) grant funded by the GIST in 2020.

\section*{References}
\bibliography{perc}

\end{document}